\documentclass[aps,eqsecnum,amsmath,onecolumn,notitlepage,nofootinbib]{revtex4-1}
\usepackage[utf8]{inputenc}
\usepackage{indentfirst}
\usepackage{amssymb}
\usepackage{amsmath,latexsym}
\usepackage{graphicx}
\usepackage{caption}
\usepackage{subcaption}
\usepackage{float}
\usepackage[margin=1.0in]{geometry}
\usepackage[mathscr]{euscript}
\graphicspath{{images/}}

\newcommand{\beq}{\begin{equation}}
\newcommand{\eeq}{\end{equation}}
\newcommand{\bea}{\begin{eqnarray}}
\newcommand{\eea}{\end{eqnarray}}
\def\beqs#1\eeqs{\beq\begin{split} #1 \end{split}\eeq}

\def\pd#1#2{\frac{\partial #1}{\partial #2}}

\usepackage[colorlinks=true,backref=false, linktocpage=true,
citecolor=blue,urlcolor=blue,linkcolor=blue,pdfpagemode=UseOutlines]{hyperref}

\hypersetup{%
  bookmarksnumbered=true,
  pdftitle = {},
  pdfsubject = {},
  pdfauthor = {},
  pdfkeywords = {}
}

\usepackage{microtype}

\newcommand{\eq}[1]{eq.~(\ref{#1})}

\DeclareMathOperator{\im}{Im}
\DeclareMathOperator{\re}{Re}

\begin{document}

\title{Fast Estimator of jacobians in Monte Carlo Integration on Lefschetz Thimbles}
\author{Andrei Alexandru}
\email{aalexan@gwu.edu}
\affiliation{Department of Physics\\George Washington University\\
Washington, DC 20052}
\author{G\"{o}k\c{c}e Ba\c{s}ar}
\email{gbasar@umd.edu}

\author{ Paulo F. Bedaque }
\email{bedaque@umd.edu}
\author{Gregory W. Ridgway}
\email{gregridgway@gmail.com}
\author{Neill C. Warrington}
\email{ncwarrin@umd.edu}
\affiliation{Department of Physics \\
University of Maryland\\College Park, MD 20742}
\date{\today}


\newpage

 \begin{abstract}
A solution to the sign problem  is the so-called ``Lefschetz thimble approach" where
the domain of integration for field variables in the path integral is deformed from the real axis 
to a sub-manifold in the complex space. For properly chosen sub-manifolds (``thimbles") 
the sign problem disappears or is drastically alleviated. The parametrization of the thimble by real coordinates require the calculation of a  jacobian with a computational cost of order 
$\mathcal{O}(V^3)$, where $V$ is proportional to the spacetime volume.
In this note we propose two estimators for this jacobian with a computational cost of order  $\mathcal{O}(V)$. We discuss analytically the regimes where we expect the estimator to work and show numerical examples in two different models.
 \end{abstract}

\maketitle

\section{Introduction}
Lattice regularized path integrals provide a means for analyzing strongly coupled systems where perturbation theory breaks down. These high dimensional integrals are most efficiently computed through Monte Carlo methods but these methods hinge on the fact that $e^{-S}/\mathcal{Z}$ is a positive definite probability distribution ($S$ is the action and $\mathcal{Z}$ the partition function of the system). Unfortunately this is not the case when the action $S$ is complex  as it is for many systems of interest. Systems with a complex action include QCD, or most models, at finite density and the Hubbard model away from half filling. This inability to consider $ e^{-S}/\mathcal{Z}$ as a probability distribution leads to the notorious ``sign problem". A geometric solution to the sign problem recently put forward is the thimble approach~\cite{Cristoforetti:2012su}. In this scheme, the path integral is extended to complex fields and the original action is analytically continued to a holomorphic function of complex fields. Picard-Lefschetz theory shows that the original integral over real fields is equal to a sum of integrals, each of which is computed over a submanifold in the complex space called a thimble. These manifolds can be chosen in such a away that the imaginary part of the action $S_{I}$ is (piecewise) constant, so the previously problematic oscillatory part of the action factors out of the partition function, leaving a positive definite probability distribution  that can be used to sample the thimble. Every implementation of this idea of a quantum field theory thus far is very computationally expensive since they all involve, for one reason or another,  the transport of a basis of vectors from point to point along the thimble. In~\cite{Cristoforetti:2013wha,Fujii:2013sra}, the transport of a basis is necessary in order to either make proposal in the Monte Carlo steps or a step in the Langevin evolution that actually lie on the thimble (a nontrivial task since the surface is curved). In the method used in~\cite{Alexandru:2015xva} one parametrizes the thimble by points in the tangent space to the critical point. The jacobian of this parametrization needs to be computed and it can be thought of as the volume spanned by the vectors obtained by transporting a unit basis to a given point on the thimble. Regardless of the reason, the computational cost of transporting a basis of the tangent space to the thimble from one point to another scales as $\mathcal{O}(V^3)$ (where $V$ is proportional to the spacetime volume), which consume the overwhelming majority of the computation resources and quickly  renders the calculation unfeasible expensive as the spacetime volume increases. Therefore, it would be beneficial to have an alternative to this procedure. In the present note, we present two such estimators of the jacobian with a computational cost of the order $\mathcal{O}(V)$  which tracks the exact jacobian with remarkable precision. 
We motivate the estimators theoretically and demonstrate through examples that these estimators can be used in Metropolis updates with the difference between the estimators and the true jacobian included by reweigthing. Two different models are used for this purpose. A fermionic model in $0+1$ dimensions and the relativistic Bose gas model in $3+1$ dimensions at finite density in a small lattice.

\section{Thimbles and the Contraction Method}
%
A critical point $z_c$ is a point in complex space where the gradient of the action vanishes:
\beq
\left. \frac{\partial S}{\partial z_i} \right |_{z_c}=0,
\eeq where $i=1,\ldots,N$.
A thimble $\mathcal{T}$  associated to the critical point $z_c$ is the union of all curves that satisfy steepest descent flow equations
\begin{equation} \label{flow}
\frac{dz_{i}}{d\tau} = -\overline{\frac{\partial S}{\partial z_i}}
\end{equation}
 approaching $z_c$ asymptotically. Here
 the bar denotes complex conjugation and $\tau$ is an auxiliary time parameter along a trajectory. Expanding the flow \eq{flow} into its real and imaginary parts,

\begin{equation} \label{SIconserved}
\begin{split}
& \frac{dz_i^R}{d\tau} = -\frac{\partial   S_R }{\partial z_i^R} = \frac{\partial   S_I }{\partial z_i^I}, \\
& \frac{dz_i^I}{d\tau} = -\frac{\partial   S_R }{\partial z_i^I} = -\frac{\partial   S_I }{\partial z_i^R},
\end{split}
\end{equation}
we find that the flow is the gradient flow of the real part of the action $  S_R $ and, at the same time, the hamiltonian flow of the ``hamiltonian"  $  S_I $. That means that the flow seeks smaller values of $  S_R $ but  conserves $  S_I $. As the thimble is defined as the set of curves that satisfy a first order differential equation, any point on a thimble $\mathcal{T}$ lies on one and only one curve that satisfies \ref{flow} and the asymptotic boundary conditions. This fact is exploited in the sampling algorithm we will use.

Near $z_c$, the action is well approximated by $S(z)=S(z_c)+\frac{1}{2} z_iH_{ij} z_j$ where the hessian $H_{i j}=\frac{\partial^2 S}{\partial z_{i} \partial z_{j}}(z_c)$ is a symmetric, but not necessarily real matrix. Expanding the quadratic action in terms of its real and imaginary parts, one finds
\begin{equation} \label{super_size}
 \frac{d}{dt}\begin{bmatrix}
          z_R  \\
          z_I   
    \end{bmatrix}=-
    \begin{bmatrix}
        H_R & -H_I \\
         -H_I & -H_R 
    \end{bmatrix}
    \begin{bmatrix}
          z_R  \\
          z_I    
    \end{bmatrix}
    =
  -  \mathbb{H}
    \begin{bmatrix}
          z_R  \\
          z_I    
    \end{bmatrix}.
\end{equation}
The reason for separating the system in its real and imaginary is that the matrix appearing on \eq{super_size} is now real and symmetric, so it affords an orthonormal basis of eigenvectors with real eigenvalues. The non-zero eigenvalues of this matrix appear in pairs $(\lambda, -\lambda)$\footnote{We will assume here that $H$ is not degenerate, that is, that no eigenvalue vanishes.}; the directions tangent to the thimble correspond to the eigenvectors with a positive eigenvalue and has, therefore, $N$ (real) dimensions.

The contraction algorithm~\cite{Alexandru:2015xva} is a Metropolis style Monte Carlo integration. 
The difficulty with applying any Markov chain or similar approaches to thimble integration  is that field configurations lying  on the thimble  must be proposed in an accept/reject step. This is not an obvious task as the geometry of the thimble is not a priori known.
To understand how the contraction algorithm handles this geometric constraint, note the following: first, under the flow, any point on the thimble will at some $\tau$ arrive arbitrarily close to the critical point. Second, since the action always increases  as the field configuration  moves away from the critical point, asymptotically distant fields are not likely to contribute significantly to any observable. Therefore, incurring only exponentially small errors, a finite (but arbitrarily  large) subset of the thimble can be sampled to compute observables. The first fact tells us that after a finite amount of flow $T$, the entire relevant subset is mapped arbitrarily close to the critical point of the thimble. Since the flowed points are in one-to-one correspondence with the original region of the thimble, the flow defines a map between the thimble and the tangent space of the critical point. Near enough to the critical point the thimble can be approximated by its tangent space. Thus, any point in the thimble (or, at least, in the regions of the thimble with substantial support in the path integral) is mapped by the flow into a point near the critical point and, consequently, into a point on the tangent space. Alternatively, a point in the tangent space near the critical point is mapped by the {\it reversed} flow to a point approaching the thimble as the flow time grows. This observation can then be used to establish a coordinate system on the thimble. Every point $z$ of the thimble is parametrized by a point $\tilde z$ on the tangent space by the flow by a fixed time $T$:
\beq
\tilde{z}\mapsto z(\tau=T)\quad\text{where}\quad
\frac{dz_{i}}{d\tau} = +\overline{\frac{\partial S}{\partial z_i}}
\quad\text{and}\quad
z(0)=\tilde z \,.
\eeq
Note the plus sign in the equation above due to fact that the (reverse) flow begins near the 
critical point and flows outwards. 
The point $\tilde z$ close to the critical point, in its turn, can be written as a (real) linear 
combination of the $N$ ``complex eigenvectors"  satisfying 
$\overline{H \rho^{(a)}}=\lambda^{(a)} \rho^{(a)}$ 
\footnote{
The $N$ N-dimensional complex eigenvectors $ \rho^{(a)}$  can be found from the $N$ 2N-dimensional real eigenvectors $( \rho^{(a)R},  \rho^{(a) I})$ of $\mathbb{H}$  by combining their real and imaginary parts as $ \rho =  \rho^R + i \rho^I$.
}:
\beq
\tilde z_i = \sum_{a=1}^N\ c_a  \rho_i^{(a)}\,.
\eeq 
The utility of the parametrization $z=z(\tilde z)$ rests on the fact that Monte Carlo updates can be
easily constrained to the thimble by proposing changes that keep the coefficients $c_i$ real. 
In fact, a change in $c_i$, and consequently, to $\tilde z_i$, leads to a change in the point  $z$ reached by the (reverse) flow after a time $T$. The larger the value of $T$ the closer to the critical point $\tilde z$ will be and more justified we are in identifying a point on the tangent space to a point in the gaussian region of the thimble.

The points on the thimble can then be sampled by sampling the variables $\tilde z$ (or  $c_i$) through, for instance,  a Metropolis algorithm. In terms of $\tilde z$ we have,  in each thimble
\beq \label{reparam}
\langle O\rangle = \frac{1}{Z}\int_{\mathcal{T}} d z \, e^{-S(z)}O(z)
=
\frac{1}{Z}\int_{\mathcal{G}} d \tilde{z} \, e^{-S(z(\tilde{z}))}  O(z(\tilde{z}))
\det\left(\frac{\partial z}{\partial \tilde{z}}\right)
=
\frac{1}{Z}\int_{\mathcal{G}} d \tilde{z} \, e^{-S_\text{eff}(\tilde{z})} O(\tilde{z}) 
\det\left(\frac{\partial z}{\partial \tilde{z}}\right)
\eeq
where $S_\text{eff}(\tilde{z})=S(z(\tilde{z}))-\text{ln}(\text{det}\left(\frac{\partial z}{\partial \tilde{z}}\right)(\tilde{z}))$, $O(\tilde{z})=O(z(\tilde{z}))$ and the set $\mathcal{G}$ is the subset of the tangent space of $z_c$ to which the relevant region of $\mathcal{T}$ is mapped under the flow. The jacobian $\det J(T)=\text{det}\left(\frac{\partial z}{\partial \tilde{z}}\right)$ 
encodes how volumes stretch and twist between the tangent space of the critical point $z_c$ and a point on the thimble. It is a complex quantity depending on the geometry of the thimble and on the particular parametrization we chose. In order to compute $\det J(T)$, 
we can take an orthonormal 
basis of the tangent space like, for instance, $ \rho^{(a)}$, and evolve it by the (reverse) flow from $\tilde z$ to $z$. The equation governing the evolution of a vector by the flow can be found considering the evolution of two infinitesimally close points  $z_1$ and $z_2$ connected by a tangent vector $v=z_1-z_2$. The flow of these points define the evolution of the tangent vector which is governed by the equation 
\beq
\frac{d v_i}{d\tau} = \overline{H_{ij}(z(\tau)) v_j} \quad\text{where}\quad H_{ij}(z)\equiv\frac{\partial^2 S(z)}{\partial z_i \partial z_j} \,.
\eeq 
The evolved basis defines a volume in complex space whose determinant is $\det J(T)$. In summary, $J$ can be found by solving
\beqs \label{eq:J}
\frac{dJ}{dt} &= \overline{H J},
\ J(0) =R\,, \\
\frac{dz_i}{d\tau} &= \overline{\frac{\partial S}{\partial z_i}},\ z(0)=\tilde z,
\eeqs
where the matrix $H(z(\tau))$ is evaluated along the trajectory $z(\tau)$. 
The initial condition for jacobian matrix $J$ is the matrix $R$ that has the vector $\rho_i$ as
the columns, that is $R_{ij}\equiv\rho_j^{(i)}$. Note that because the vectors $\rho_i$ are
orthogonal not only in terms of the real scalar product, which is true because the real
$\mathbb{H}$ matrix is symmetric, but also in terms of complex scalar 
product\footnote{This extra condition in true because $\im \rho_i^\dagger \rho_j=0$, due
to fact that the vectors $i\rho_i$ are also eigenvectors of $\mathbb{H}$ orthogonal on $\rho_j$.}.
If we normalize these vectors, the matrix $R$ is unitary and its determinant is a phase.

The cost of computing the jacobian can be split into two parts. One, is the diagonalization of the Hessian $\mathbb{H}(z_\text{cr})$ defining the tangent space which is $\mathcal{O}(V^3)$ operation. This step, however, needs to be done only once in the beginning of the calculation. The cost of solving the system of differential equation in~\eq{eq:J} is dominated by the matrix multiplication and is also generically of order $\mathcal{O}(V^3)$. In some special models, where the action is local (typically purely bosonic models) the Hessian is a sparse matrix and the computational cost of this step is $\mathcal{O}(V^2)$.



\section{The Estimators}


In this section we propose the use of two different estimators of $J$. The first one, $W_1$ is given by
\beq\label{wrongian_1}
\log J \approx \log W_1 = \int^T_0 dt'\ \sum_a { \rho}^{(a) \dagger} \overline{H}(t')\bar{ \rho}^{(a)},
\eeq where $\rho^{(a)}$ are the complex eigenvectors of the matrix $H(z_c)=\partial^2 S(z_c)/\partial z_i \partial z_j$ with positive eigenvalues. In the gaussian region, where the action is well approximated by its quadratic part
\beq
S(z) \approx S(z_c) + \frac{1}{2} (z-z_c)^TH(z_c)(z-z_c),
\eeq $W_1$ agrees with $J$. In order to see that we notice that \eq{eq:J} has the formal solution
\beq
J(t) = R + t  \overline{H}  \overline{R} + \frac{t^2}{2!}  \overline{H} H R + \frac{t^3}{3!}  \overline{H} H  \overline{H}   \overline{R} + \cdots ,
\eeq 
Indeed,
\beqs
\frac{dJ(t)}{dt} &=  \overline{H}  \overline{R} + t  \overline{H} H R + \frac{t^2}{2!}  \overline{H} H  \overline{H}  \overline{R} + \cdots \\
&=  \overline{H} \overline{\left[ R + t   \overline{H}  \overline{R} + \frac{t^2}{2!}   \overline{H} H  R + \cdots\right]} =  \overline{H} \bar J\,.
\eeqs
But, since $\overline{ H \rho^{(a)}} = \lambda^{(a)} \rho^{(a)}$, $\overline{H R} = \Lambda R$ where $\Lambda=\text{diag}(\lambda^{(1)}, \lambda^{(2)}, \cdots)$.  Thus,

 \beq
 J(T) = R + T \Lambda R + \frac{T^2}{2!} \Lambda^2 R + \cdots = e^{T \Lambda}R
 \eeq 
 and
 \beq
 \text{det}J(T) = \text{det}(e^{T \Lambda}) \text{det} (R) = e^{T \sum_a \lambda^{(a)}}
 \det R\,.
 \eeq 
 On the other  hand, again in the quadratic approximation, we have
 \beq
 \log W_1 
 = \int^T_0 dt'\ \sum_a { \rho}^{(a)\dagger} \overline{H}{ \bar\rho}^{(a) }
 =T \sum_a { \rho}^{(a)\dagger}{ \rho}^{(a)} \lambda^{(a)}
=T \sum_a  \lambda^{(a)} \,,
 \eeq 
 and we see that these expressions only differ by $\det R$ which is a fixed phase.

The second estimator of $J$ we propose is
\beq \label{wrongian_2}
W_2 = \exp{\int_0^T dt' \text{Tr}   \overline{H}(t')}.
\eeq A theoretical justification for its use can be found by noticing that the time-ordered exponential
\beq
P \exp{\int_0^T dt'  \text{Tr} \overline{H}(t')}
\eeq is the solution of
\begin{equation} \label{wrongian_1}
\frac{dJ}{dt} = \overline{H}J,
\end{equation} which, up to the complex conjugation of $J$, is the same as \eq{eq:J}. Presumably $W_2$ is a good estimator of $J$ in situations when $J$ is mostly real.

%
%
%
%

The  cost of computing $W_1$ and $W_2$ is somewhat dependent on the model and it will be discussed below; in practice they are computationally much cheaper than $\det J$. 
Their usefulness relies on that and on the fact that they
track the actual value of $\det J$ well.  Note, they can be used for Monte Carlo runs as long as the difference between $\det J$ and $W_{1,2}$ is reweighted with the observable. More precisely, we can write the expectation value of an observable as:
\beqs\label{eq:reweight}
\langle \mathcal{O}\rangle &=
\frac1Z
{\int d\tilde z\ \mathcal{O} \det J e^{-  S_R }}
 \\
&=
\frac1Z
{\int d\tilde z\ \mathcal{O} e^{i \im\log\det J} e^{-  S_\text{eff} }}
 \\
&=
\frac1Z
{\int d\tilde z\ \mathcal{O} e^{i \im\log\det J} e^{-  S_\text{eff} +S'_\text{eff} } e^{-  S'_\text{eff} }}
 \\
&=
\frac1{Z'}
{\int d\tilde z\ \mathcal{O} e^{i \im\log\det J} e^{-  \Delta S } e^{-  S'_\text{eff} }}
\Bigg/
\frac1{Z'}
{\int d\tilde z\ e^{i \im\log\det J} e^{-  \Delta S } e^{-  S'_\text{eff} }} \\
&=
{\langle  \mathcal{O} e^{i \im\log\det J} e^{-  \Delta S } \rangle_{S'_\text{eff}}  }
\big/
{\langle e^{i \im\log\det J} e^{-  \Delta S }  \rangle_{S'_\text{eff}}  },
\eeqs
where $S_\text{eff}=  S_R -\re\log\det J$,  $S'_\text{eff}=  S_R -\re\log\det W_1$ or 
$S'_\text{eff}=  S_R -\re\log\det W_2$ and $\Delta S = S_\text{eff}-S'_\text{eff}$. 
$Z$ and $Z'$ are the partition functions for $\det J\times \exp(-S_R)$ and $\exp(-S'_\text{eff})$
respectively.
It is evident from the formula above that the efficacy of the method relies on on the quantity 
$\exp(i \im\log\det J)\exp(-  \Delta S )$ to fluctuate little from one field configuration to the next. 
Some of this fluctuation comes from $\exp(i \im\log\det J) $ (called the ``residual phase") and is intrinsic to the geometry of the thimble and is unrelated to the estimator. The other 
factor, $\exp(-  \Delta S )$, arises from the use of the estimator.
In the next section we present tests of this idea in specific models.

\section{Results}

The first model we will consider is a $0+1$ dimensional version of the Thirring model with (continuous) action given by
\begin{equation}\label{eq:S-fermions}
L_\text{Th.}= \bar \chi \left (\gamma^0 {d \over dt} + m +\mu \gamma^0\right) \chi+ {g^2 \over 2}\left(\bar \chi \gamma^0 \chi \right)^2 \,,
\end{equation}
where $\chi$ is a two component, time dependent spinor and $\gamma^0$ is a Pauli matrix. After discretizing it (using staggered fermions) and introducing a bosonic auxiliary variable $\phi$ we arrive at the lattice model
\beq
Z=\left[\prod_{t=1}^{N} \int_{0}^{2\pi} {d \hat\phi_t\over 2\pi}\right] \det D e^{-{1\over 2\hat g^2} \sum_{t=1}^N (1-\cos \hat\phi_t)}\equiv\left[\prod_{t=1}^{N} \int_{0}^{2\pi} {d \hat\phi_t\over 2\pi}\right] e^{-S[\hat\phi]}, \label{eq:Z-lattice}
\eeq
where the effective action and the explicit form of the discretized Dirac matrix are
\beqs
S[\hat\phi]&={1\over 2\hat g^2} \sum_{t=1}^N (1-\cos \hat\phi_t)-\log\det D[\hat\phi] \,, \\
D_{t,t^\prime}[\hat\phi]&={1\over 2}\left(e^{\hat\mu+i \phi_t}\delta_{t+1,t^\prime}-
e^{-\hat\mu-i \hat\phi_{t^\prime}}\delta_{t-1,t^\prime}+
e^{-\hat\mu-i \hat\phi_{t^\prime}}\delta_{t,1}\delta_{t^\prime, N} -
e^{\hat\mu+i \hat\phi_{t}}\delta_{t,N}\delta_{t^\prime, 1}\right)+\hat m\,\delta_{t,t^\prime}\,. 
\eeqs
Here $N=\beta/a$ is an even number that denotes the number of lattice sites related to 
the inverse temperature of the system $\beta$, and all the dimensionful quantities, 
$m$, $g^2$, $\mu$ are rendered dimensionless  by multiplying with appropriate 
powers of the lattice spacing: $\hat m=ma$, $\hat\mu = \mu a$, $\hat g^2=g^2 a$, $\hat\phi=a\phi$. This model has a sign problem at finite $\mu$ that can be severe at large enough $g^2$ and $\mu$. However, it is easily solved analytically and has been used as a testing ground of different approaches to the sign problem \cite{Pawlowski:2013pje}. It has also been studied with the Lefshetz thimble approach where the need to include multi thimble contributions were detected \cite{Fujii:2015bua,Fujii:2015vha,Alexandru:2015xva}   and approaches to include them were discussed \cite{Alexandru:2015sua}. Here we will not discuss the agreement between the Monte Carlo calculations and the exact result. Instead, we focus on the feasibility to use one of the estimators instead of the actual jacobian in the manner described by \eq{eq:reweight}.

\begin{figure}
\includegraphics[width=.6\textwidth]{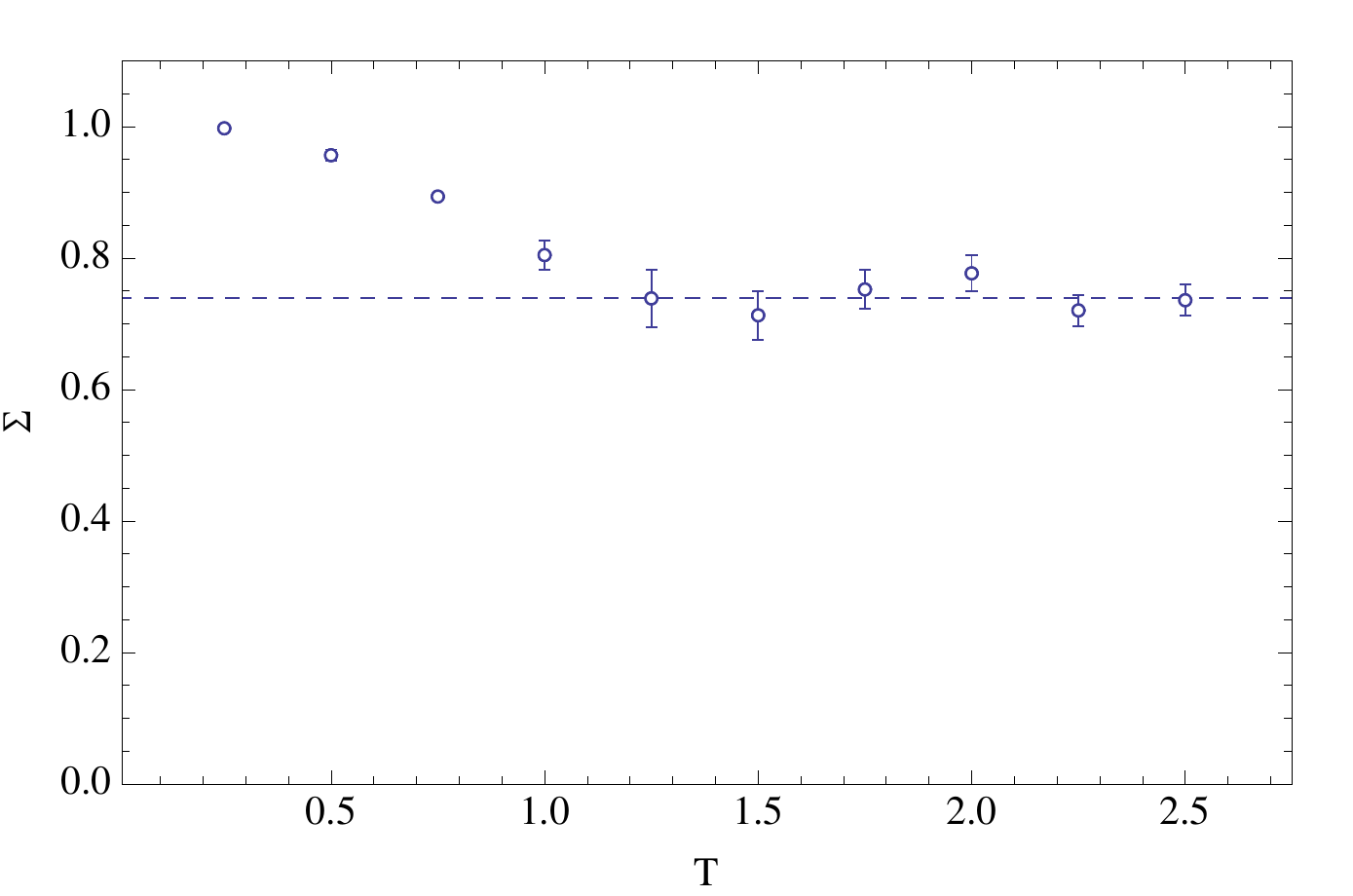}
\caption{Statistical power for the parameter set $N=8, \hat m=1, \hat\mu=1, \hat g^2=1/2$ as a function of flow time. The line indicates the average value of the last 6 points.}
  \label{fig:thirring8}
\end{figure}

For that we repeat the calculations in \cite{Alexandru:2015xva} by now using the estimator $W_1$. If the estimator is accurate, the ratio $w=e^{-\Delta S} /\langle e^{-\Delta S} \rangle$ fluctuates little and the weight of every configuration to $\langle \mathcal{O}  \rangle$ is similar. On the other hand, if only a few configurations have a large value of $w$ and dominate the reweighting, large statistical errors are expected. This observation is made quantitative by the statistical power $\Sigma$ defined as
\beq
\Sigma=
\frac{1}{\mathcal{N}}
\frac
{\langle w \rangle_{S'_\text{eff}}  }
{\langle w^2 \rangle_{S'_\text{eff}}  },
\eeq 
where $\mathcal{N}$ is the number of configurations used in the averages. If all configurations have the same statistical weights in the reweight procedure the statistical power is $1$; on the other extreme, if only one configuration has weight one while all the other configurations have zero weight the statistical power attains its smallest possible value $1/\mathcal{N}$.

The critical point with the smallest (real part of the) action is a constant field configuration $\phi_t = (\phi, \cdots, \phi)$ for a certain value of $\phi$. The tangent space to the associated thimble (at the critical point) is purely real. Consequently, $W_1=W_2$ in this case. The cost of computing the estimator $W_1$ in this model is dominated by the cost of computing the trace of the hessian matrix at every point of the flow trajectory connecting $\tilde z$ to $z$ and is of order $\mathcal{O}(V)$, compared to the $\mathcal{O}(V^3)$ cost of computing $\det J$ 
\footnote{Of course, in this simple model an explicit formula for the determinant can be found and, in this sense, the cost of computing $\det J$ is not $\mathcal{O}(V^3)$. This is a special feature of this $0+1$ dimensional model and does not generalize to higher dimensions while the estimates above do. }.

In order to test the effectiveness of the jacobian estimators we use $W_1$ in the Monte Carlo runs (or, equivalently, the accept/reject step is perfomed using $S'_\text{eff}$). The configurations thus obtained are then reweighted as shown in \eq{eq:reweight} and the statistical power of the reweighting is computed.

To integrate the flow equations we use an adaptive integrator~\cite{Cash:1990:VOR:79505.79507}, which uses a fourth and fifth order Runge-Kutta methods to evaluate
the next point along the flow and also to estimate the errors. The step size
is adjusted up or down to keep the errors at the desired level. The observable used to monitor 
the error dictates the size of the integration steps. We monitor both the position
along the flow, $z(t)$, and  the jacobian (or its estimator) and use
the largest error to adjust the step size. It turns out that the jacobian is more sensitive
to the integration errors and it determines the step size. It is important to note that
when integrating the estimator along the flow, the integrator does not use the
same steps as used in the calculation of the jacobian. This has two important consequences. First, the estimator is less
sensitive to numerical errors, probably because its calculation does not require
an LU decomposition as is necessary for the jacobian. This makes the integration
run even faster since it requires fewer steps. On the other hand, the calculation
of  $z(T)$ produces slightly different results when integrating the jacobian
or the estimator. The observables are evaluated using the value of $z(T)$ 
produced by the more precise jacobian integration. To account for the fact that
$z'(T)$, the result of estimator integration, differs slightly from the exact $z(T)$,
we need to reweight using $S'_\text{eff}(z'(T))=S_R(z'(T))-\re \log \det W_1$. Note
that $z'(T)$ can be made arbitrarily close to $z(T)$ by tightening the integrator
errors. From a numerical perspective, it is more profitable to use a looser and
faster integrator, as long as $S'_\text{eff}(z'(T))$ differs little from $S'_\text{eff}(z(T))$.
The final results has no bias, but the error bars might increase due to the
additional reweighting due to the difference between $S'_\text{eff}(z'(T))$ and
 $S'_\text{eff}(z(T))$.

In ~Fig.~\ref{fig:thirring8} we show the statistical power  with parameters $N=8$, $\hat m=\hat \mu=1$ and $\hat g^2=1/2$. The statistical power for both estimators is close to one indicating that both $W_1$ and $W_2$ are useful. 
One might naively think that the estimators are reliable for short flow times $T$ but will degrade for larger flows. After all, $\det J=W_1=W_2=1$ for $T=0$. As Fig.~\ref{fig:thirring8}  shows, the opposite is  true. This can be understood by noticing that the points sampled on the thimble are determined by the physics of the model and not by our choice of parametrization. Consider the parametrization with two different flow times $T$ and $T'$. In these two parametrizations the same point $z$ on the thimble is parametrized
by two different points $\tilde z$ and $\tilde z'$.
The path connecting $\tilde z$ to $z$ (for flow time $T$) and the path connecting  $\tilde z'$ to the same $z$ (for flow time $T'$) differ only by the path connecting $\tilde z$ to $\tilde z'$ which lies very close to the critical point. In that region, as shown above, $W_1$ is a good estimator of $\det J$. Thus, beyond a certain flow time the estimate of $\det J$ by $W_1$ does not get worse.

\begin{figure}
\begin{subfigure}{.5\textwidth}
  \centering
  \includegraphics[width=.95\textwidth]{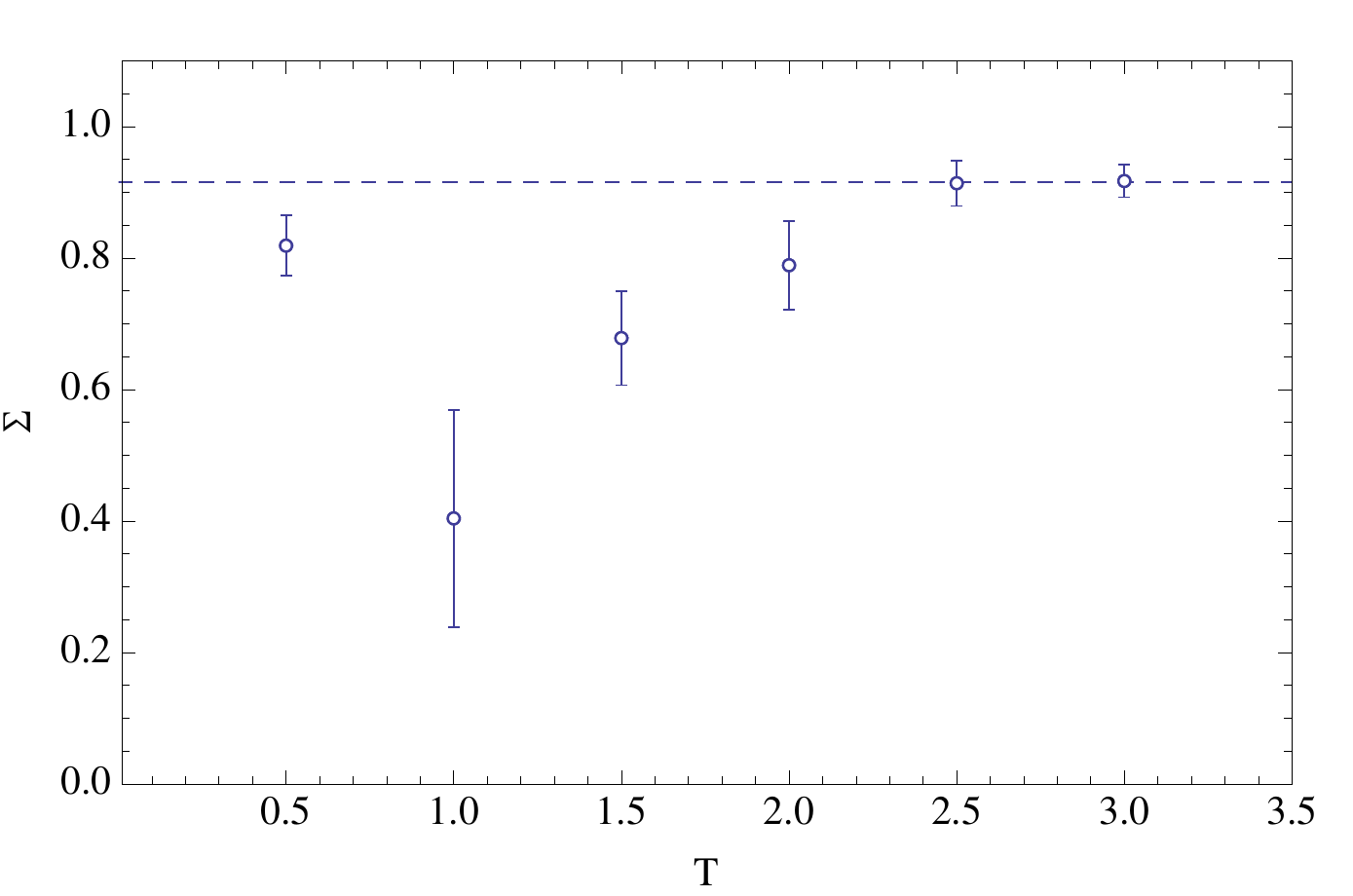}
  \caption{Low temperature limit: $N=32, \hat m=1, \hat\mu=1, \hat g^2=1/2$.}
  \label{fig:thirringcold32}
\end{subfigure}%
\begin{subfigure}{.5\textwidth}
  \centering
  \includegraphics[width=.95\textwidth]{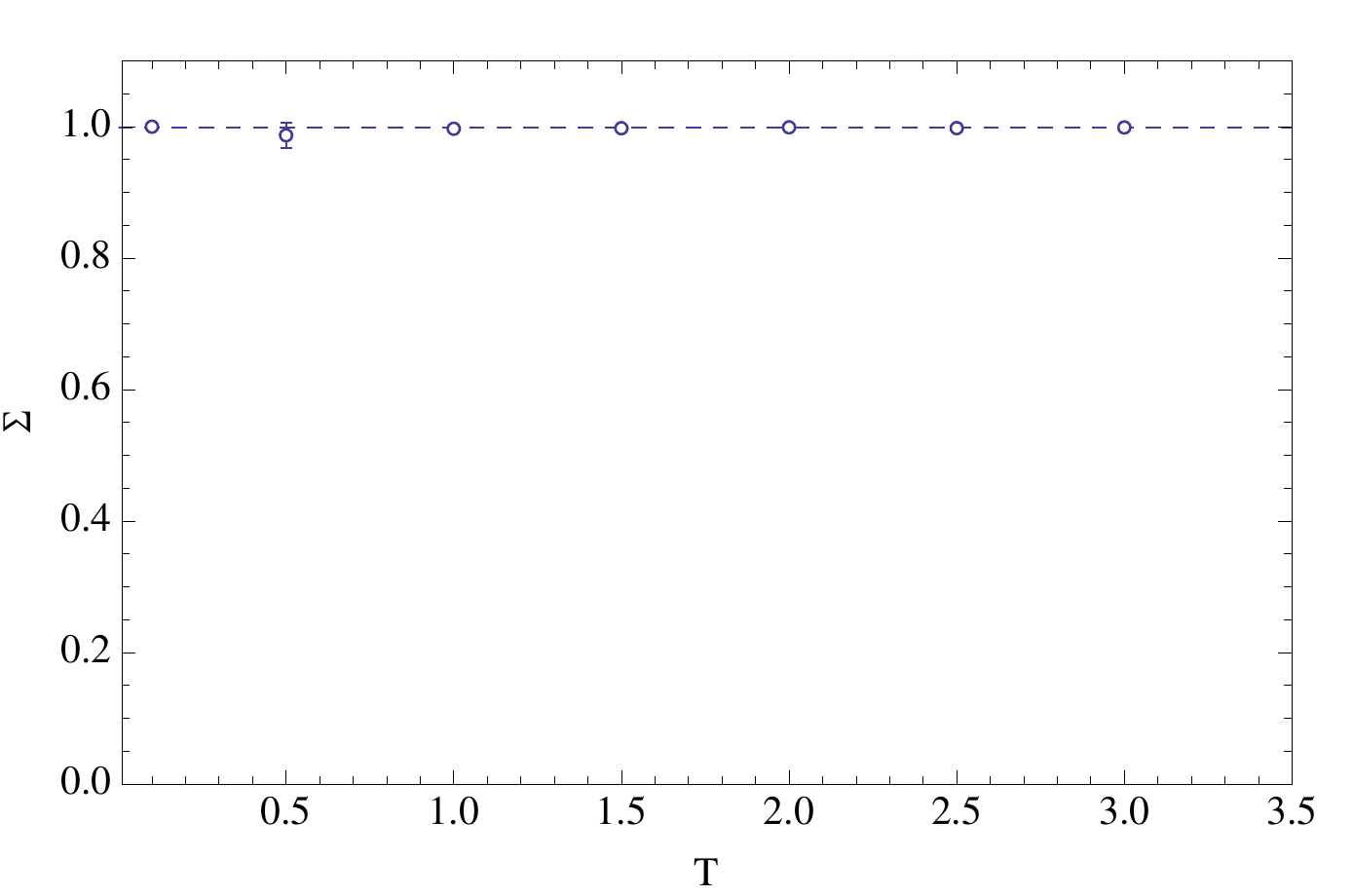}
  \caption{Continuum limit: $N=32, \hat m=1/4, \hat\mu=1/4, \hat g^2=0.322$.}
  \label{fig:thirringcontinuum32}
\end{subfigure}
\caption{Statistical power of reweighting for $N=32$.}
\end{figure}

The natural question is whether the estimators still track the jacobian for larger values of $N$. There are two ways of taking the large $N$ limit: by lowering the temperature with a fixed lattice spacing or by decreasing the lattice spacing at a fixed temperature. In the first case only the parameter $N$ changes while in the second one all parameters have to be scaled appropriately. Fig.~\ref{fig:thirringcold32} shows the statistical power for the parameters $N=32$, $\hat m=\hat \mu=1$ and $\hat g^2=1/2$, corresponding to a temperature 4 times smaller than in Fig.~\ref{fig:thirring8}. Fig.~\ref{fig:thirringcontinuum32} shows instead the statistical power for the parameter set $N=32$, $\hat m=\hat \mu=1/4$ and $\hat g^2=0.322$
\footnote{The parameter scaling we use in the continuum limit is explained in \cite{Alexandru:2015sua}.}
 corresponding to a continuum limit extrapolation of the parameter set used in Fig~\ref{fig:thirring8}.  As explained in \cite{Fujii:2015vha,Alexandru:2015sua}, for flows times larger than $T>0.5$ they correspond to the contribution of one thimble only while the full result receives non-negligible contributions from other thimbles. 
 The drop in statistical power between $T=0.5$ and $T=1.0$ is probably related
 to the fact that calculations with small flow times correspond to an integration over  a manifold very different from the one thimble sampled in the $T\rightarrow\infty$ limit (this point is discussed extensively in~\cite{Alexandru:2015sua}).
 As such we should concentrate on the larger flows and, for those, the statistical power is high.
 

%
%
%


The second model we consider is a system of relativistic spin-0 bosons with a finite chemical potential. The lattice action is given by
\beq
S=\sum_x \bigg[ \left(4+\frac{m^2}2 \right) \phi^a_x\phi^a_{x} - \sum_{\nu=1}^3 \phi_x^a \phi_{x+\hat\nu}^a 
-\cosh \mu \phi_x^a \phi_{x+\hat0}^a 
+ i\sinh\mu \epsilon_{ab} \phi_x^a \phi_{x+\hat0}^b + \frac\lambda4 (\phi_x^a\phi_x^a)^2 
-h(\phi_x^1 + \phi_x^2) \bigg] \,, 
\eeq 
where $\epsilon$ is the antisymmetric tensor with $\epsilon_{12}=1$.
The Hessian is given by
\beqs
\pd{}{\phi_x^b}\pd{}{\phi_y^c} S &= (8+m^2)\delta_{xy}\delta^{bc}
-\sum_{\nu=1}^3 \left(\delta_{x+\hat\nu,y}+\delta_{x-\hat\nu,y} \right)\delta^{bc}
-\cosh\mu \left(\delta_{x+\hat0,y}+\delta_{x-\hat0,y} \right)\delta^{bc} \\
&\qquad   +i\sinh\mu \left(\delta_{x+\hat0,y}-\delta_{x-\hat0,y} \right)\epsilon_{bc}
+\lambda \delta_{xy} \left[ (\phi_x^a\phi_x^a)\delta^{bc}+2\phi_x^b\phi_x^c \right] \,.
\eeqs
The computation of $W_1$ for this action starts by the computation (only once) of the basis $ \rho^{(a)}$. For the present hessian $H(\phi)$ 
we have
\beq
H(\phi)_{x,y}^{bc} = (H_0)_{x,y}^{bc} + \lambda
\delta_{xy}
\left[(\phi_x^a\phi_x^a)\delta^{bc}+2\phi_x^b\phi_x^c-(\phi\to\phi_\text{cr})\right]\,,
\eeq 
where $H_0=H(\phi_{cr})$. In order to compute $W_1$ we define
\beq
\Delta(\phi)\equiv
\sum_{i,x,y,b,c} \overline{( \rho^{(i)}_0)_{bx}}   \delta_{xy} \left[ (\phi_x^a\phi_x^a)\delta^{bc}+2(\phi_x^b\phi_x^c) 
\right] \overline{( \rho^{(i)}_0)_{cy}} = \sum_x \sum_{bc} {\cal R}_x^{bc} \left[ (\phi_x^a\phi_x^a)\delta^{bc}+2(\phi_x^b\phi_x^c) \right] \,,
\eeq where $ \rho^{(i)}_0$ are the ``complex eigenvector" of $H_0$ and
\beq
{\cal R}_x^{bc} \equiv \sum_i\overline{( \rho^{(i)}_0)_{bx}} \  \overline{( \rho^{(i)}_0)_{cx}} \,,
\eeq
a vector of $2\times2$ matrices that can be precomputed. We have
\beq
\sum_a \rho^{(a)\dagger} \overline H(\phi)  \bar\rho^{(a)} = 
\sum_a \rho^{(a)\dagger} \overline H_0 \bar\rho^{(a)}  + \overline{\Delta(\phi)} - 
\overline{\Delta(\phi_\text{cr})} = \left(\sum_i \lambda_i - \overline{\Delta(\phi_\text{cr})}\right) +
\overline{\Delta(\phi)} \,,
\eeq
where $\lambda_i$ are the positive eigenvalues of $\mathbb{H}$ at the critical point.
The quantity in the parenthesis is computed once at the beginning of the calculation and
the only term that has to be computed at every step is $\overline{\Delta(\phi)}$. 
The calculation of $ \rho^{(a)\dagger}\overline H \bar\rho^{(a)}$
is then a ${\cal O}(V)$ operation. 
Similarly, the computation of $W_2$ involves only the trace of $H(\phi)$ which is a $\mathcal{O}(V)$ operation. 

\begin{figure}
  \centering
  \includegraphics[width=.6\textwidth]{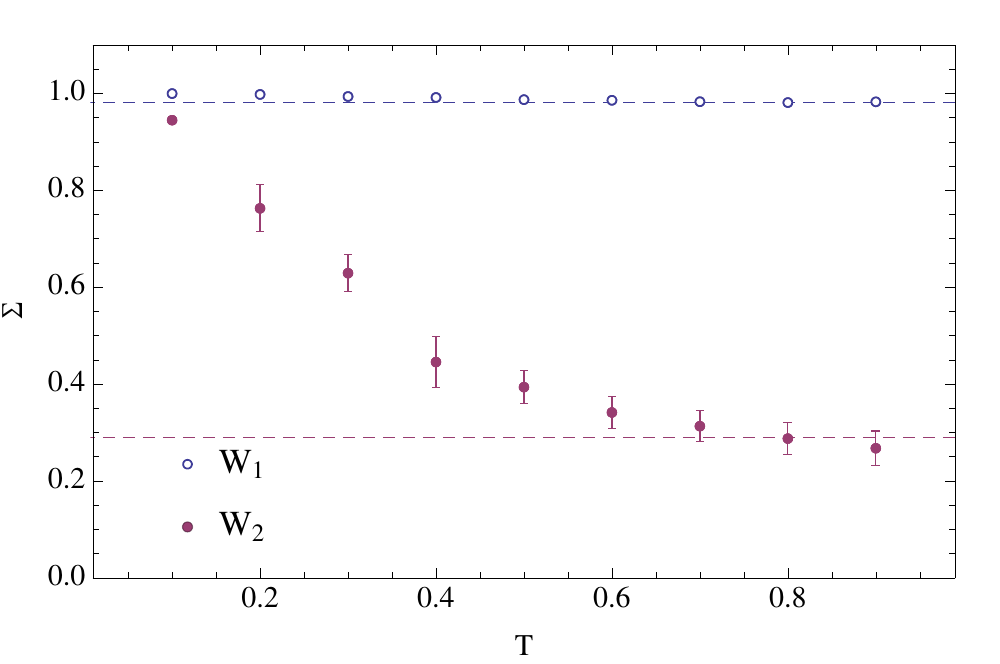}
  \caption{
  Statistical power of the bosonic model  on a $4^4$ lattice with parameters, $\hat m=1, \hat\mu=1.3, \lambda = 1$ and $h=0.03+ i 0.003$ for the two estimators $W_1$ and $W_2$. }
  \label{fig:phi4}
\end{figure}

The statistical power of both estimators, $W_1$ and $W_2$ are shown in Fig~\ref{fig:phi4} 
for a $4^4$ lattice and parameter $m=1$, $\mu=1.3$ and $h=0.03+i 0.003$.  The symmetry breaking parameter $h$ was chosen complex to avoid the appearance of the Stokes phenomenon (a thimble connecting two critical points) which violates the decomposition of the original integral into a sum of thimbles. 

The main feature seen in Fig~\ref{fig:phi4} is  that the statistical power of the $W_1$ reweighting remains very close to one at all values of flow time tested. The estimator $W_2$ also tracks the jacobian reasonably well and has the advantage of being easier to implement since it does not require the knowledge of the eigenvalues $\rho_0^{(a)}$. The hit on the statistical value of configurations generated using $W_2$ is more than compensated by the gain in speed during the Monte Carlo runs. In fact, for the parameters we explored the Monte Carlo runs are a factor of $2000$ faster using $W_2$ than using the jacobian $\det J$ while the statistical power is reduced by only a factor $\approx 3$. Furthermore, the statistical power does seem to saturate at large flow times and not asymptote to zero. 

\section{Conclusions}
We presented in this note two fast estimators for the jacobian that arises when parametrizing a thimble as used in the contraction algorithm. We showed that these estimators yield statistically robust data sets (in the sense of statistical power) for a $0+1$ dimensional fermion model and a $3+1$ dimensional bosonic model in a small lattice. The computational costs of the estimators are $\mathcal{O}(V)$ ($V$ is the spacetime volume) while the original jacobian requires a number of order $\mathcal{O}(V^3)$ of operations.
Even in the small volumes considered  in our experiments ($4^4$ in the bosonic model) this is an improvement by a factor of approximately $2000$ in efficiency and it is essential for the practical use of the algorithm.
 Furthermore, we found that these estimators do not lose their utility in the limit of large flows. This is an extremely useful trait, especially for large lattices which require large flows to tame the sign problem. The authors are confident that the estimators described in this paper will prove useful for a variety of models in the future.

\acknowledgements

A.A. is supported in part by the National Science Foundation CAREER grant PHY-1151648. 
A.A. gratefully acknowledges the hospitality 
of the Physics Department at the University of Maryland where part of this work was 
carried out.
G.B., P.B., G.R and N.C.W.  are supported by U.S. Department of Energy under Contract No. DE-FG02-93ER-40762.

\medskip

\bibliographystyle{JHEP}
\bibliography{thimbles}

\end{document}